\documentclass[%
 aip,apl,showplace,
rsi,%
 amsmath,amssymb,
 reprint,%
]{revtex4-1}

\usepackage{graphicx}
\usepackage{dcolumn}
\usepackage{bm}

\begin{document}

\preprint{AIP/123-QED}

\title[Sample title]{Bandpass transmission spectra of a whispering-gallery microcavity coupled to an ultrathin fiber}

\author{Fuchuan Lei}
 \affiliation{Light-Matter Interactions Unit, Okinawa Institute of Science and Technology Graduate University, Onna, Okinawa 904-0495, Japan}
\author{Rafino M. J. Murphy}%
\affiliation{Light-Matter Interactions Unit, Okinawa Institute of Science and Technology Graduate University, Onna, Okinawa 904-0495, Japan}%
\author{Jonathon M. Ward}%
\affiliation{Light-Matter Interactions Unit, Okinawa Institute of Science and Technology Graduate University, Onna, Okinawa 904-0495, Japan}%
\author{Yong Yang}%
\affiliation{Light-Matter Interactions Unit, Okinawa Institute of Science and Technology Graduate University, Onna, Okinawa 904-0495, Japan}%
\affiliation{National Engineering Laboratory for Fiber Optics Sensing Technology, Wuhan University of Technology, Wuhan, 430070, China}

\author{S\'ile Nic Chormaic}%
 \email{sile.nicchormaic@oist.jp}
\affiliation{Light-Matter Interactions Unit, Okinawa Institute of Science and Technology Graduate University, Onna, Okinawa 904-0495, Japan}%

\date{\today}

\begin{abstract}
Tapered fibers with diameters ranging from 1-4 $\mu$m are widely used to excite the whispering-gallery (WG) modes of microcavities. Typically, the transmission spectrum of a WG cavity coupled to a waveguide around a resonance assumes a Lorentzian dip morphology due to resonant absorption of the light within the cavity. In this paper, we demonstrate that the transmission spectra of a WG cavity coupled with an ultrathin fiber (500-700nm) may exhibit both Lorentzian dips and peaks, depending on the gap between the fiber and the microcavity. By considering the large scattering loss of off-resonant light from the fiber within the coupling region, this phenomenon can be attributed to partially resonant light bypassing the lossy scattering region via WG modes, allowing it to be coupled both to and from the cavity, thence manifesting as Lorentzian peaks within the transmission spectra, which implies the system could be implemented within a bandpass filter framework.
\end{abstract}

\maketitle

In recent years, whispering-gallery (WG) mode microcavities have attracted considerable interest for both fundamental research and applications\citep{ilchenko2006optical,peng2014parity,yang2016high,sumetsky2004whispering,ward2009trapping, jiang2016chip,yang2016tunable,ward2016glass,wang2016packaged}. The main advantage of WG cavities are their ultra-high quality (\textit{Q}) factors and small mode volumes, inherent attributes that allow for the effective trapping of light both spatially and temporally. Consequently, the use of WG resonators makes it possible to significantly enhance light-matter interactions and, as a result, they have been used for studying strong coupling between a single atom and a cavity mode\cite{aoki2006observation}, ultra-low threshold Raman lasing\cite{spillane2002ultralow}, parametric oscillation\citep{kippenberg2004kerr,farnesi2014optical}, frequency comb generation\cite{yang2016four,del2007optical,kato2017transverse} and optomechanical effects \cite{aspelmeyer2014cavity,madugani2015optomechanical,lei2015three}, etc.
When performing studies of this nature, it is imperative that light is coupled in and out of the microcavity with high efficiency. Among various coupling schemes, the tapered fiber side-coupling technique stands out from the rest and it is predominantly used due to its inherent ultra-low loss and easy manipulation\cite{knight1997phase}, especially for silica microcavities. 

Unlike the Fabry-P\'erot cavity, the typical transmission spectrum of the WG microcavity side-coupled to a waveguide exhibits a Lorentzian dip around a resonant frequency. The transmitted light is a superposition of two components: one directly passing through the waveguide and another coming from the cavity. A $\pi$ phase difference between these two components results in destructive interference\cite{haus1984waves}.
These Lorentzian dips have been explored for use in narrow band-stop filters. However, in order to realize bandpass function, an add-drop configuration is needed\cite{rokhsari2004ultralow,wang2016packaged}, requiring the inclusion of additional coupling devices.
In this paper, we observe that the transmission spectra of a WG cavity side-coupled with a single tapered fiber may exhibit Lorentzian peak behavior when the diameter of the tapered fiber is smaller ($\sim$500-700 nm) than conventional values ($\sim$1-4 $\mu$m).
Considering that the scattering loss due to the tapered fiber is large when a microcavity is in close proximity to it, we can attribute the phenomenon to partially resonant light bypassing the strong scattering region mediated by cavity modes, while off-resonant light  experiences significant scattering, resulting in appreciable transmission loss. \\

\begin{figure}[b]
\centerline{\includegraphics[width=7.5cm]{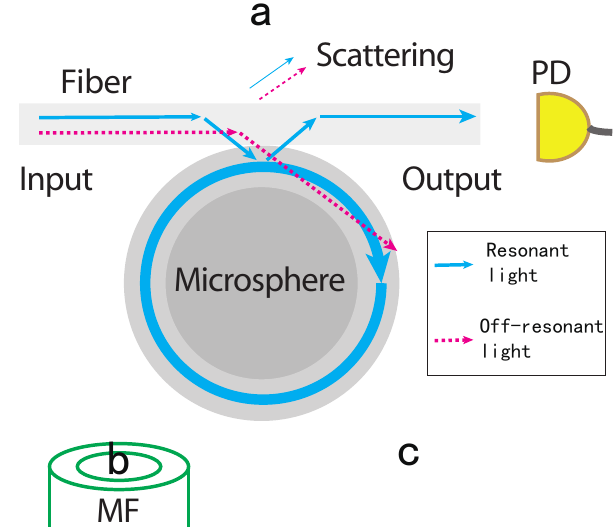}}
\caption{Schematic of the experimental setup. A tapered fiber is side-coupled to a microsphere for transmission spectra measurements. Aside from direct monitoring of the transmitted light by a photodetector (PD), a multimode fiber (MMF) is mounted at different positions (a,b,c) to collect the scattered light from the tapered fiber-microsphere coupled system. The blue solid line and red dashed line represents the resonant light and off-resonant light, respectively.}
\label{figure0}
\end{figure}

\begin{figure}
\centerline{\includegraphics[width=8.5cm]{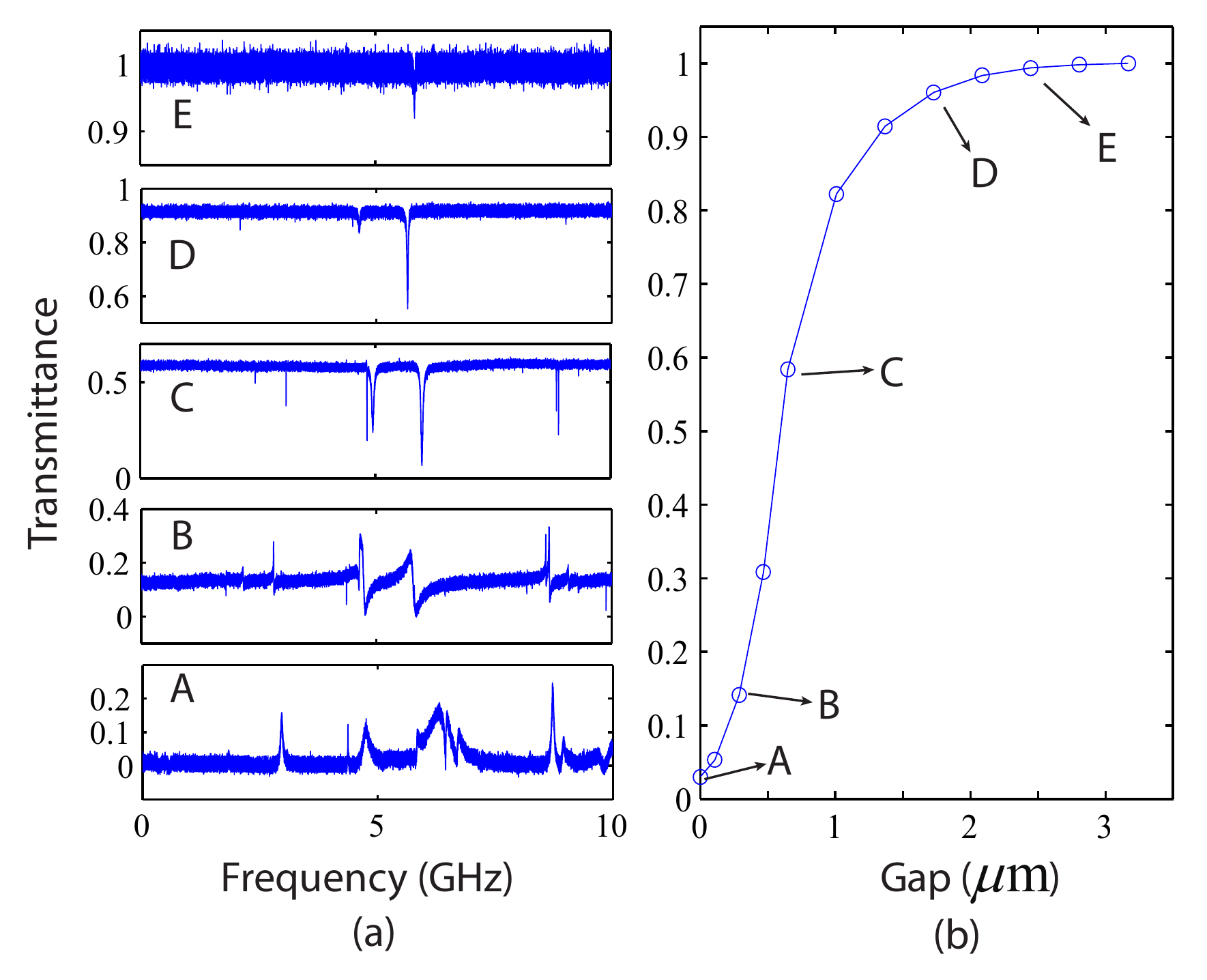}}
\caption{Normalized transmission spectra (a) and corresponding transmittance of off-resonant light (b) for varying taper-microsphere gaps. From A to E, the gap increases from 0 to 2.4 $\rm \mu m$.
For both (a) and (b) the y-axis corresponds to tranmittance, which has been normalized with respect to the maximum value observed in transmission spectrum E.} \label{figure1}
\end{figure}

The schematic setup of our experiment is shown in Fig. \ref{figure0}. The WG microcavities in our experiment were fixed-stem, silica microspheres with diameters ranging from 40-150 $\mu$m, fabricated from single-mode fiber (SMF-28, Thorlabs). We first prepared a sharp-tipped glass filament using a focused $\rm CO_{2}$ laser beam and subsequently reheated the tip. Surface tension within the molten silica caused the tip to assume a spherical morphology. The tapered fiber was fabricated by heating a strand of SMF-28 with a ceramic heater and simultaneously stretching it. In this work, the waist diameter of the tapered fiber we used was 500-700 nm measured by scanning electron microscopy (SEM). Light was sent through one end of the taper to couple both in and out of the WG microsphere in the $1550~{\rm nm}$ band. The transmitted light was detected by a photodetector (PD). Aside from the PD, a multimode fiber (MMF) was mounted near the system in the equatorial plane perpendicular to the stem of the sphere. The MMF was used to collect the scattered light and was connected to another PD (not shown in Fig. \ref{figure0}).  
\begin{figure}[h]
\centerline{\includegraphics[width=8.5cm]{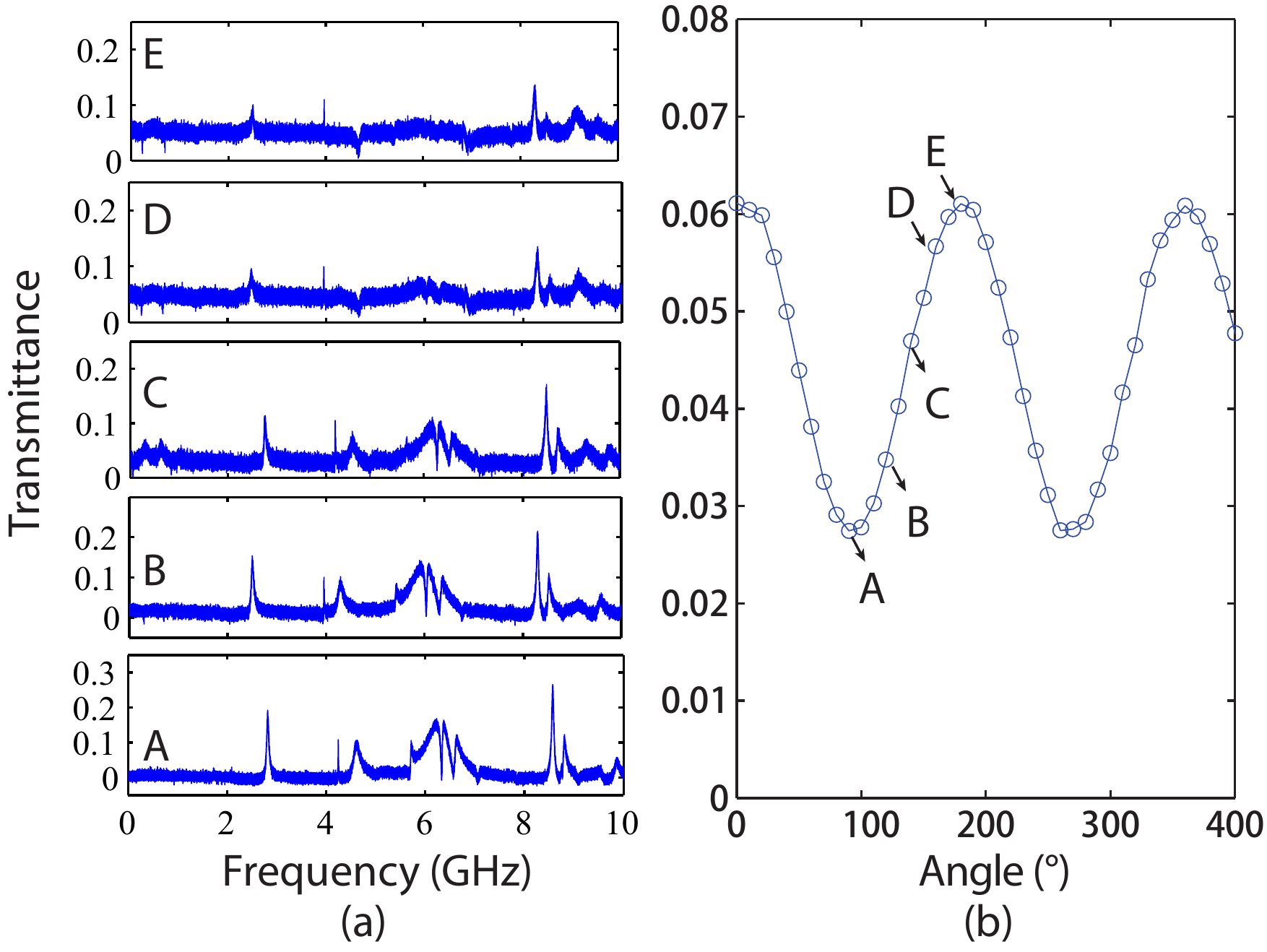}}
\caption{Transmission spectra (a) and corresponding transmittance of off-resonant light (b) with varying input polarization direction. From A to E, the polarization direction changes by $90^\circ$. The polarization was incremented in steps of $10^\circ$ for a net change of $400^\circ$.
}\label{figure2}
\end{figure}

\begin{figure}[t]
\centerline{\includegraphics[width=7cm]{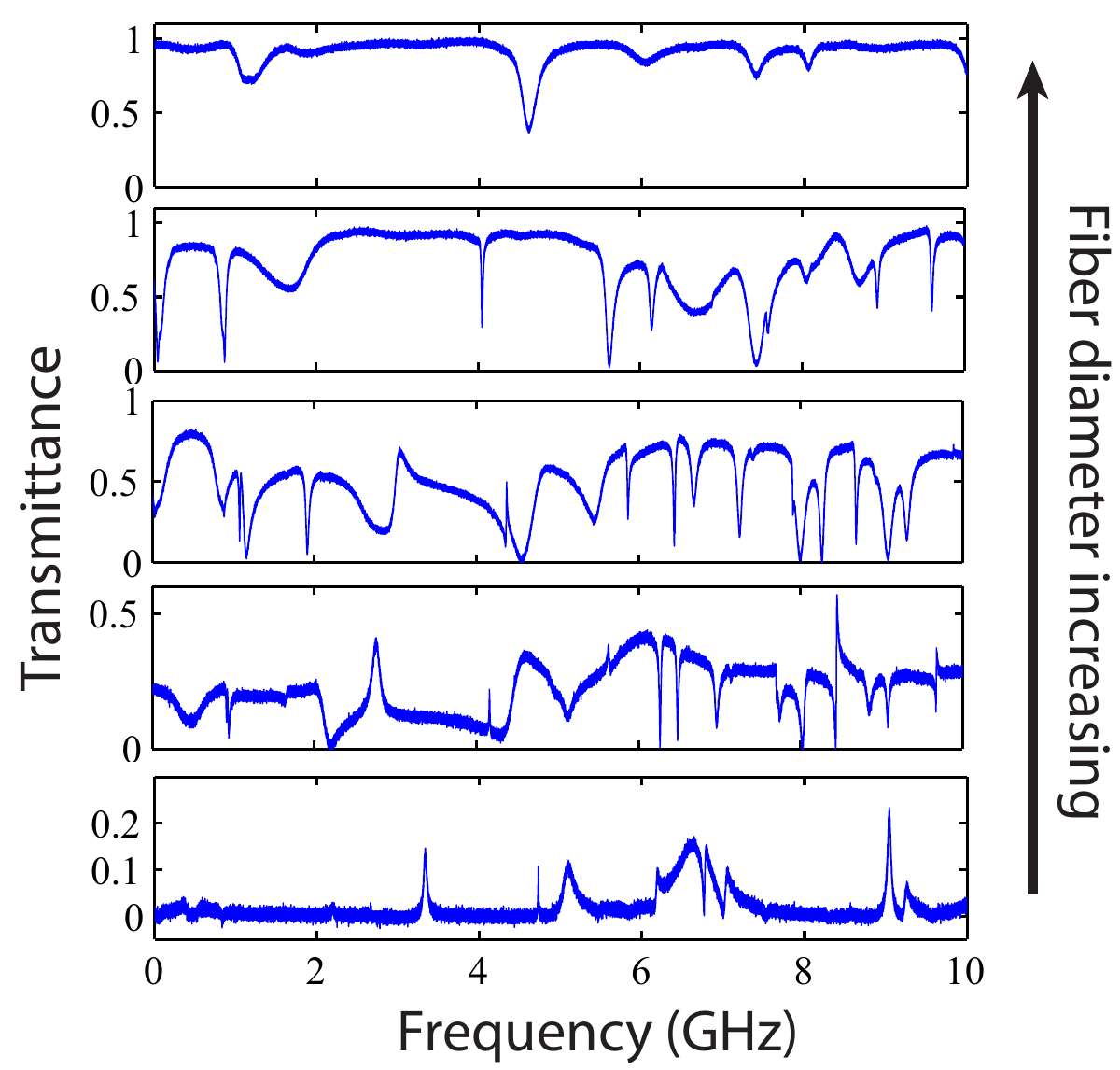}}
\caption{Transmission spectra for contact coupling with different fiber diameters. From bottom to top, the tapered fiber was translated to increase its diameter at the coupling point. According to the displacement, the diameter was estimated to be 600 nm, 800nm, 1$\mu$m, 1.5$\mu$m and 3 $\mu$m, respectively. }\label{fig4}
\end{figure}

Fig. \ref{figure1}(a) shows the evolution of the transmission spectrum as the tapered fiber-microsphere gap was altered. From A to E, the gap increased from 0 to 2.4 $\mu$m. While the gap was decreased gradually (E $ \rightarrow $ C) more and more WG modes were excited. However, some Lorentzian dips disappeared and some Lorentzian peaks emerged as the gap was decremented further (C$\rightarrow$A). At the same time, the normalized transmittance of off-resonant light decreased from unity to nearly zero, as shown in Fig. \ref{figure1}(b). This indicates that scattering loss due to the microsphere's close proximity to the nanofiber was no longer negligible in contrast to that seen in previous reports (less than 5\%)\cite{cai2000observation} -- especially when the gap between the two approaches zero. Since the tapered nanofiber diameter was of sub-wavelength dimensions, the fiber mode possessed an evanescent component that extended significantly into the free space surrounding the tapered region, e.g. the evanescent field of the $\rm HE_{11}$ mode at 1550 nm band can extend about 3 $\mu$m from the surface for a 600 nm diameter fiber. As shown in Fig. \ref{figure1}(a), almost all (97$\%$) the off-resonant light was lost when the microsphere was in contact with the ultrathin fiber, while a reduced amount of the resonant light survived (more than 20$\%$). It is worth noting that the linewidth of one particular peak is less than 10 MHz (indicative of a high-\textit{Q} factor of $\sim10^7$) even while the resonator is in contact with the ultrathin fiber.

Fig. \ref{figure2}(a) depicts a series of transmission spectra for varying linear polarization orientations. The direction of polarization was altered by rotating a half-wave plate prior to the the linearly polarized light coupling into the ultrathin fiber. Though it is evident that the transmission spectra change with the direction of polarization, many modes still retain their Lorentzian peak morphology. We also note that the transmittance of the off-resonant light varies sinusoidally with the input polarization, shown in Fig. \ref{figure2}(b). Considering that the ultrathin fiber can only support the fundamental mode $\rm HE_{11}$ and its  field distribution is determined by the polarization of the input light\cite{tong2011subwavelength}, it follows by extension that the scattering strength should be polarization dependent.

Fig. \ref{fig4} shows a series of contact-coupling transmission spectra for tapered fibers of varying thickness. From bottom to top, the fiber diameter is increased from 600 nm to about 3 $\mu$m. As the fiber diameter increases, the spatial extent of the surrounding evanescent field diminishes, resulting in a decrease in scattering losses and  the transmission of off-resonance light increases from nearly zero to unity. Consequently, as the fiber diameter is increased the Lorentzian peaks disappear, and the characteristic resonance Lorentzian dips are reinstated. \\

\begin{figure}[t]
\centerline{\includegraphics[width=7.5cm]{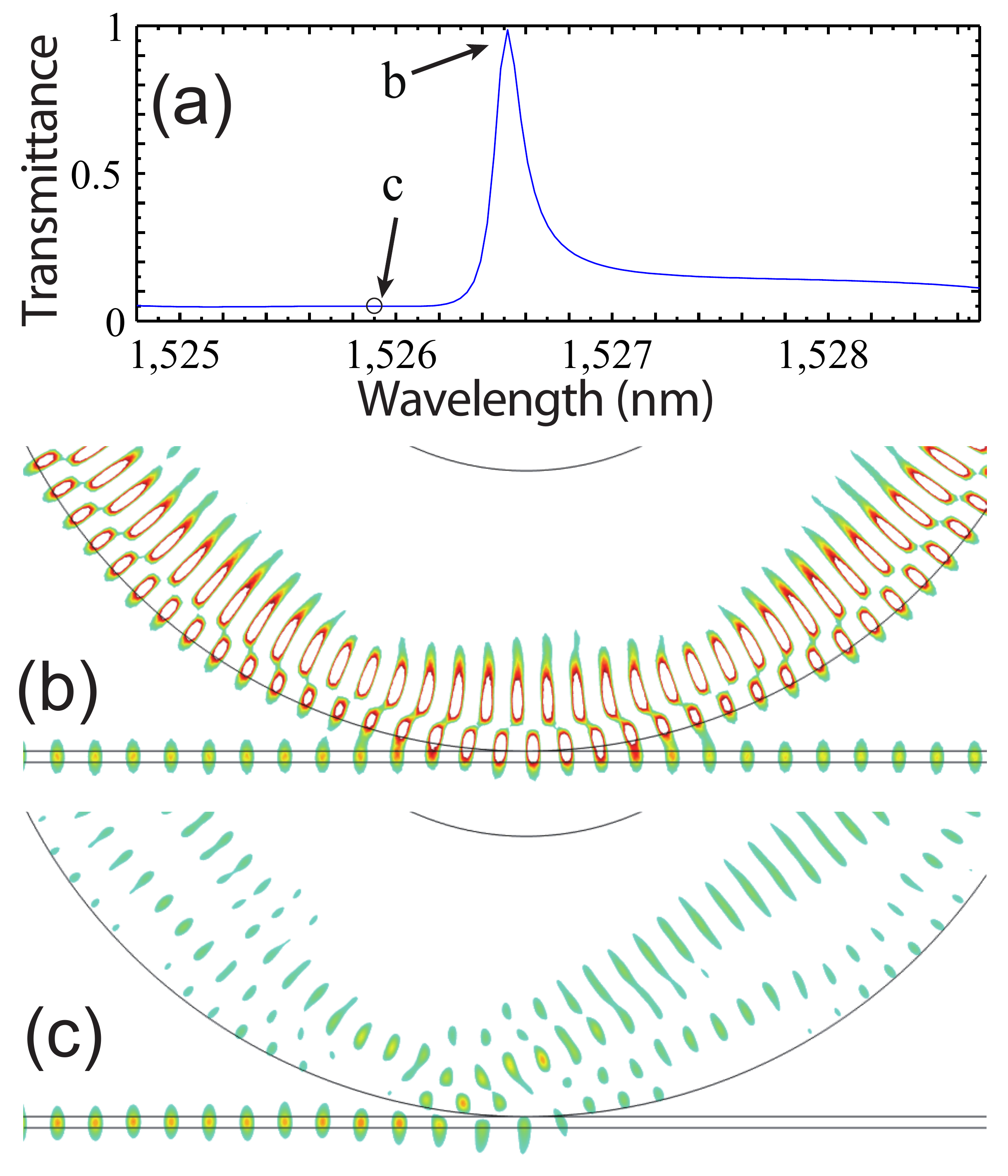}}
\caption{FEM simulation results. (a) The transmission spectrum; The distribution of the electrical field for (b) resonant case and (c) off-resonant case.}\label{figure5}
\end{figure}

In order to better illustrate the scattering loss and how it manifests in the transmission spectra, we performed finite element method (FEM) numerical simulations by using COMSOL, the results of which are given in Fig. \ref{figure5}. As a proof-of-principle, a 2-D model was constructed for simplicity. The thickness of the waveguide used in simulations was 400~nm with a refractive index of 1.35 simulating the nanofiber used in the experiment. The microsphere is simplified using a micro-ring cavity model with a refractive index of 1.45, thickness of 10 $\mu$m and outer diameter of 40 $\mu$m. The transmission spectrum from 1524.8~nm to 1528.7~nm is calculated and shown in Fig. \ref{figure5}(a). Similar to the observed spectra, e.g., see the bottom of Fig. \ref{figure1}(a), there is a peak at 1526.5~nm, corresponding to a second order radial mode. The electric field distribution perpendicular to the taper of the aforementioned mode is shown in Fig. \ref{figure5}(b), while Fig. \ref{figure5}(c) corresponds to the off-resonant light case (1525.9~nm). 

\begin{figure}[t]
\centerline{\includegraphics[width=7.5cm]{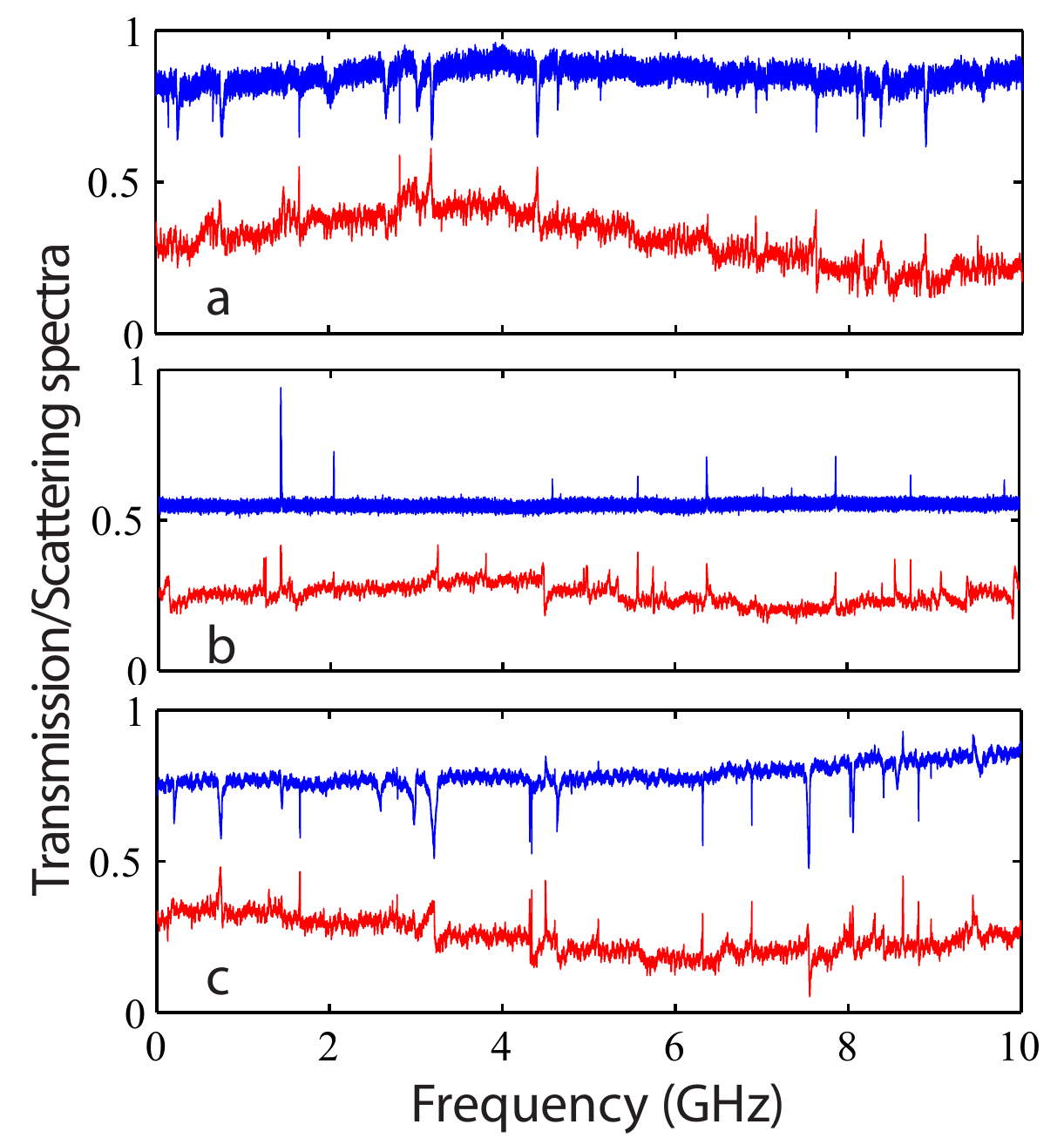}}
\caption{Spectra of the transmission through the ultrathin fiber (red, bottom) and the scattering spectra from the multi-mode fiber (blue, top). For ease of comparison, the spectra are not normalized but shifted relative to each other. a, b, and c correspond to the multi-mode fiber being placed at positions a, b, and c positions in Fig.\ref{figure0}. }\label{fig6}
\end{figure}

From the calculated field distribution in Fig. \ref{figure5}(b) and (c), the scattering caused by the sphere can be explained as follows. The effective mode index of the fiber taper is quite small compared to that of the microsphere since its diameter is far less than the operating wavelength (e.g., the effective mode index of the fundamental mode $\rm HE_{11}$ at 1550 nm band for a 600 nm diameter fiber is only 1.025). Light is almost completely refracted into the sphere if the fiber is close to it, thus it no longer propagates directly along the fiber. Nevertheless, the light can be partially coupled back to the fiber if it is resonant with a cavity mode. The light can have a very high intensity and circulate along the periphery of the sphere (confined by total internal reflection), as plotted in Fig. \ref{figure0}. Conversely, the short lifetime of off-resonant light within the cavity does not allow a strong field to build up, thus decreasing the amount of light that can be coupled back into the fiber.

To confirm our hypothesis, we monitored the optical field distribution of the coupling system by introducing a multimode fiber (MMF) to collect the scattered light. The MMF was placed at positions a, b, c (see Fig. \ref{figure0}),  the transmission spectrum at each location was recorded  and these are shown in Fig. \ref{fig6} (blue, upper curves). For ease of comparison, we plot both the spectra of the fiber taper output (red, lower curves) and the MMF transmission spectra, none of which have been normalized, but whose transmission values have been shifted relative to each other. For this experiment, we used an erbium-doped fiber amplifier (EDFA) to amplify the input signal in order to augment the scattered signal strength. EDFA power fluctuations resulted in the non-flat transmission baseline evident in Fig. \ref{fig6}. When the MMF was located at positions a and c, we observed both strong scattering signals and numerous dips in the spectra. This implies that the collected off-resonant scattered light dominates over the resonant light, which was confined in the cavity and coupled out to the ultrathin fiber. At position b, only resonant light losses were collected by the MMF, resulting in the peaks evident in the spectrum. This leads us to the conclusion that the amount of scattered off-resonant light is much weaker at position b than at positions a and c. By extension, it also demonstrates that some WG modes are excited by the ultrathin fiber. 
It should be noted that some peaks in the ultrathin fiber transmission spectra coincide with some of the peaks that are observed in the MMF spectrum, indicating that those peaks are indeed from resonant light within the cavity modes. In fiber-coupled WG microcavity systems, there may be several other mechanisms based on mode interference that lead to the peak formation in the transmission spectra, such as electromagnetically induced transparency\cite{yang2015coupled,chiba2005fano} and Fano resonances\cite{li2011experimental,miao2016dynamic,shang2017experimental}. However, the aforementioned mechanisms do not explain the simultaneous occurrence of peaks in both the ultrathin fiber and MMF spectra.

In summary, we have studied the transmission spectra of WG microcavities with respect to tapered ultrathin fiber coupling systems. Within this coupling framework, the off-resonant scattering losses from the narrow tapered fiber region while in close proximity to the microsphere have been found to be significant, and play a crucial role in the formation of the Lorentzian peaks observed in the transmission spectra. The existence of these Lorentzian peaks in the experimental setup suggests that the system, in principle, could be used as a band pass filter. Additionally, this work portrays an alternative means of controlling the behavior of transmission spectra by manipulating the amount of scattering loss. This may be useful when striving for high experimental precision and may have advantages for sensing applications.

This work was supported by the Okinawa Institute of Science and Technology Graduate University. The authors wish to thank J. Du for measuring the fiber diameters. 

\nocite{*}
\bibliography{aipsamp}
\end{document}